\documentclass[twocolumn,showpacs,preprintnumbers,amsmath,amssymb]{revtex4}
\usepackage{makeidx}
\usepackage{graphicx}% Include figure files
\usepackage{dcolumn}% Align table columns on decimal point
\usepackage{bm}% bold math

\begin{document}

 \title
{Funnelling effect  in networks}
\author {Parongama Sen}
% \email{psphy@caluniv.ac.in}
\affiliation{
Department of Physics, University of Calcutta,92 Acharya Prafulla Chandra Road,
Calcutta 700009, India.\\
}

%\address
%  {
%         Department of Physics, University of Calcutta,
%             92 Acharya Prafulla Chandra Road, Kolkata 700009, India. \\
%             parongama@vsnl.net,subinay@cubmb.ernet.in\\
%  }

\begin{abstract}
Funnelling effect, in the context of searching on networks,  precisely 
indicates that the search takes place through a few specific nodes. 
We define the funnelling capacity $f$ of a node as the fraction of successful 
dynamic paths through it with a fixed target. 
The distribution $D(f)$ of the fraction of nodes 
with funnelling capacity $f$  shows a  power law behaviour 
in  random networks (with power law or 
stretched exponential degree distribution) for a considerable range of 
values of the parameters defining the networks.  
Specifically we study in detail $D_1=D(f=1)$, which is the quantity signifying 
 the presence of nodes through which all the dynamical paths pass through.
In scale free networks with degree distribution $P(k) \propto k^{-\gamma}$, 
$D_1$ increases linearly with $\gamma$ initially and then attains a constant value. 
It shows a power law behaviour,   $D_1 \propto N^{-\rho}$,  with the number of nodes $N$
where $\rho$ is weakly dependent on $\gamma$ for $\gamma > 2.2$.
The latter variation is also independent of the number of searches.
On stretched exponential networks with $P(k) \propto \exp{(-k^\delta)}$, 
$\rho$ is strongly dependent on $\delta$.
The funnelling distribution for a model social network, where the question 
of funnelling is most relevant, is also investigated. 
\end{abstract}
\maketitle

\section{Introduction}
Searching on networks has attracted a lot of attention recently.
In general, the problem is to send a signal to a target node from a source node.
It has been shown in some  studies on real networks that it is possible to find
short  paths
 \cite{milgram,killworth,white,dodds,adamic_search,hong,geog}  during such a searching or navigation on 
small world and scale-free networks. 
This  implies that even with only local knowledge of the network, there is a small world effect, i.e.,
the average number of steps to reach the target is $O(\log(N))$ where $N$ is the number of nodes in
the network.
 In several theoretical
studies the 
interest therefore has been to find out the scaling relation of the 
shortest searching path lengths
with the number of nodes using different searching algorithms \cite{klein,adamic1,kim,zhu,moura,watts-search,carmi,thada,clauset,sen1,sen2,sen3,boguna1}.

The notion  of the small world effect emerged from the results of  the original 
 experiments made by Milgram \cite{milgram}  
in which
letters had to be hand delivered to a specific target. Apart from the observation of 
small world effect it
was also claimed that  
the successful paths filtered through a few nodes \cite{white} and this effect was 
termed funnelling. 
In these experiments however, very few chains were completed and the results
could be less than conclusive.
In a later study  by Dodds et al \cite{dodds}, where search experiments
on email networks 
were conducted, it was concluded that
no such funnelling effect exists for social networks.

 The question of   funnelling  
 has not been adequately addressed so far in any theoretical study to the best of
our knowledge.
In fact, 
no precise quantitative definition of funnelling has been 
proposed either. 

We define the funnelling capacity of a node to be the  
  fraction of dynamic paths through  it
when the target is fixed and the source is varied. 
%This is
%in tune with the idea of funnelling in Milgram's and later experiments. 
In a realistic search, failure to reach the target has a 
considerable probability and  
searches with  a possibility  of termination  \cite{sen1,sen2} have been
studied earlier. 
Hence, with a fixed target, we define the  funnelling capacity
$f_i$ of the $i$th node as 
\begin{equation}
f_i = \frac{No ~~of~~successful~~ searches~~ through~~ the~~ ith ~~node  }{Total~~ no~~ of~~ successful~~ paths}.
\end{equation}

Defined in this way, it may seem that the funnelling capacity, averaged over all targets 
 is the  same as the
betweenness centrality \cite{between} of the node. The latter is defined as the 
fraction of shortest paths through a node  and is a much studied quantity, but it must be remembered
that  it
 is obtained from the global knowledge of
the network, and is thus a static property.
Thus these two quantities are expected to behave differently in general.
Funnelling capacity, which is  a dynamic variable will obviously  depend 
on the search algorithm.

It may also be mentioned that keeping the target
 fixed is an important criterion; 
a node in general is not expected to be part of the traffic for all
choices of source-target pairs (the target is selected randomly).  
 On the other hand if one relaxes this criterion, 
only the hubs can show the funnelling effect. 
In the  experiments of social searching also, funnelling has been 
considered by keeping the target fixed \cite {milgram,dodds}. 

%In this paper, we report our studies on the 
% distribution of the funnelling capacity 
%in different types of networks and
%for different algorithms. 
%We first focus on 
%We find  that  the behaviour of this distribution changes significantly 

%when the parameters governing the structure of the network or the searching algorithm are varied.

%

%Apparently, the funnelling effect should exhibit itself by 
%a 
%power law decay in the distribution $D(f)$ of the funnelling 
%capacity which
%we define as the fraction of nodes having funnelling capacity $f$.
In the present work, we have carried out simulations, in which,
 after generating the desired network, we fix a target and allow 
different nodes to be the source nodes.   The dynamic 
path to the target (if it exists) is then found out to calculate the funnelling capacity.
We then obtain the distribution $D(f)$, which is precisely the fraction  of nodes with
funnelling capacity $f$. 
To obtain $D(f)$, such searching processes are repeated on many networks.

Strictly speaking, $f_i$ is dependent on the target node as well,
but here we have not studied that aspect directly. Rather,  we 
expect that the dependence will be reflected in the distribution $D(f)$ itself.

The presence of funnelling effect would imply that $D_1= D(f=1)$ should be non-zero. We 
have therefore focussed our attention on this quantity and studied its behaviour as a function 
of the parameters of the network.

%The question naturally arises that what are the networks and parameters 
%used in the study. 
We have studied some random networks with given  degree distribution
as well as 
 a correlated network which can serve as a toy
model of a social network. 
The chosen degree distributions are either scale-free or stretched exponential
type and are controlled by suitable parameters.
 
In addition, we consider as a parameter $\nu$, the ratio of the
number of searches to be made (i.e., number of  sources) to the total
 number of nodes $N$. 
%This has been done for several fixed values of 
%$\nu$ in scale free networks.
Unless otherwise mentioned, the value of $\nu$ has been taken to be 0.1.

Since the degree of a node is an important quantity of a network and it is customary
to study the behaviour of quantities as a function of degree, we have also
studied the average funnelling capacity $\langle f(k) \rangle$
of nodes with degree $k$ in case of scale-free networks.  

We have used Monte Carlo simulations to study the funnelling effect by generating networks of size $\leq 5000$ and
taking averages over typically 10000 to 20000 network configurations. 

In sections II and III the results for scale-free  and stretched exponential
networks are presented respectively.
 In section IV,  funnelling distribution in the toy
model of social network has been discussed. In the last section we have summarised the results 
and drawn a few concluding remarks.

%We find that a power law behaviour upto $f \to 1$ exists when we say that
%there is a funnelling effect.
%In absence of the funnelling, there is an exponential decay of $D(f)$ as $f \to 1$.
%In fact, we find that there is a more intriguing behaviour of the 
%distribution close to $f=1$ than a simple power law when the network shows a funnelling effect.
%This will be discussed in detail later. 
%It 
%can be driven by varying the parameters defining the network
%or by altering the algorithm.

\section{Funnelling in scale free networks}

In this   section we discuss the results   for  networks 
which are constructed with a scale free  degree distribution 
but are otherwise random.
The  search algorithm which has been used is degree based; 
such  algorithms have  been considered in networks (especially for scale free networks) 
quite commonly \cite{adamic1,kim}.

We have generated random scale
free networks with degree distribution $P(k) \propto k^{-\gamma}$
 with the  exponent $\gamma$ 
lying between 2 and  3 corresponding to realistic networks.
The generation of the networks and the algorithm are described in detail in \cite{sen1}.
We allow a minimum of two links (degree) 
for each node, while
the maximum is $N^{1/2}$. 
All links are undirected and there are no multiple links
between a given pair of nodes.
We have used two degree based algorithms described in the 
following subsections. There is a general
rule that a node can receive a message only once and  searches  terminate in case 
there is no neighbour left to whom the message can be passed. 

\subsection {Highest degree search (HDS)}

Here the message is passed to a neighbour with the
highest degree $d_h$ (highest degree search or HDS).  In case of multiple neighbouring nodes with degree $d_h$,
one is selected randomly. 
However, if the target node happens to be a neighbour of a node, the
message will be conveyed to it without considering the degree of 
other neighbours.
When finding out the neighbour to whom the signal is to be passed,
the neighbours which have already received it once are  not
considered.

\begin{center}
\begin{figure}
%\vskip -1cm
\includegraphics[clip,width= 5.0cm, angle=270]{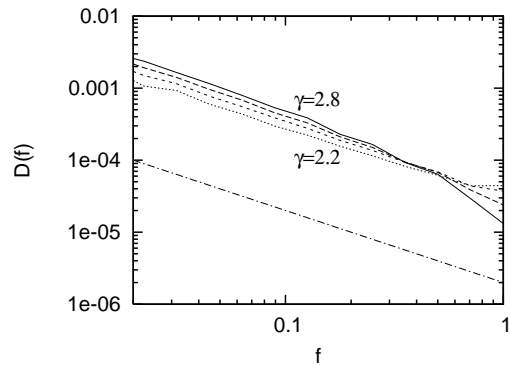}
%\includegraphics[clip,width= 5cm, angle=270]{funnmostlogbin2.eps}
%\ includegraphics{funnmost1.eps}
%\includegraphics{funnmost2.eps}
\caption{ The funnelling distribution $D(f)$ is shown for  
scale-free networks with $\gamma~=~2.8,~2.6,~2.4$ and
$2.2$ (from top to bottom) for $\nu = 0.1$. The straight line has slope equal to -1.
}

%against $\lambda$ for
%a random scale-free network for values of $k_{min} = 1,2,3 $ and $4$.
%               }
\label{funfig1}
\end{figure}
\end{center}

 With this algorithm, 
it is observed that $D(f)$ shows a power law decay against $f$ with exponent close to 1
but as     $\gamma$ increases  beyond $\gamma=\gamma^* \simeq 2.4$, the power law behaviour is observed only for 
a limited range of  $f$   
 and shows a  more rapid decay to zero as  $f \to 1$. 
The results are shown in Fig. \ref{funfig1}.
%The funnelling effect is said to exist if there are nodes through which
%most of the dynamical paths  pass for  a fixed target, i.e., there
%are a finite number of nodes with $f$ very close to one.  Hence, essentially 
%the behaviour of $D(f)$ close to $f=1$  would determine whether there is 
%a funnelling effect present in the searching or not. Thus if the power
%law behaviour continues till $f=1$, a funnelling effect is said to exist.
%The power law region of $D(f)$, which extends to $f=1$ for  
%$\gamma < \gamma^* \simeq 2.4$ can be well approximated by $D(f) \propto f^{-1}$ for all $\gamma < \gamma^*$ indicating a universal behaviour 
%below $\gamma^*$.

%\end{center}

We find another intriguing behaviour of $D(f)$ for  $\gamma$ values below 
$\gamma^*$. Here, $D(f)$ actually shows a tendency to increase for $f$ 
very close to unity. 
(By definition the maximum value of $f$ is one and therefore 
the increase in $D(f)$ cannot continue indefinitely. )
In fact, even for $\gamma > \gamma^*$, $f=1$ is a special point where
 $D(f)$ shows a significantly higher value than that at $f$ just below unity causing
 a discontinuity in $D(f)$. 
This is another reason to study the behaviour of $D_1$ more intricately.
The reason for the   discontinuity in $D(f)$ for $\gamma > \gamma^*$   is 
apparently 
due to the presence of a  few nodes through which the
searching path always passes (e.g.,    the nearest neighbours of the target).

\begin{center}
\begin{figure}
\includegraphics[clip,width= 5.0cm, angle=270]{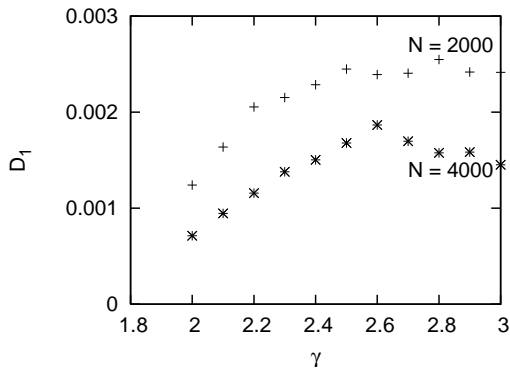}
\caption{The value of the distribution $D(f)$ at $f=1$ in scale-free networks 
is shown against $\gamma$
for two different network sizes.}
\label{funfig2}
\end{figure}
\end{center}

In order
to investigate the behaviour at $f=1$ more closely, we have plotted $D_1$ as a function 
of $\gamma$ (Fig. \ref{funfig2}).  We notice  that $D_1$ first increases linearly 
with $\gamma$ and
then tends to attain a constant value at higher $\gamma$. 

The $\gamma=2.0$ is a special point where the 
average degree shows a logarithmic divergence.
If the average degree is large, $D_1$
will naturally be small as there are many available neighbouring
nodes to pass on the signal.  
Hence the value of $D_1$ initially increases with $\gamma$. However, as
$\gamma$ increases further the number of hubs decrease and consequently $D_1$
does not increase anymore.
It can be expected 
that $D_1$ should decrease for very high values of $\gamma$, however, for $\gamma \leq 3.0$ this tendency is not  strongly evident.

\begin{center}
\begin{figure}
\includegraphics[clip,width= 5.0cm, angle=270]{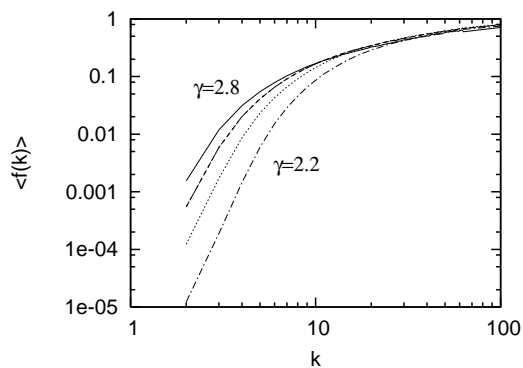}
\caption{The average funnelling 
capacity as a function of degree is shown for scale free networks with $\gamma=2.8,2.6,2.4$
and $2.2$.}
\label{funfig3}
\end{figure}
\end{center}

\vskip -0.25cm
As $\gamma$ increases, the number of nodes with large degree 
becomes less and the above observation indicates that the funnelling
capacity of nodes with less degree must increase with $\gamma$. Indeed, this is evident from  
 a plot (Fig. \ref{funfig3}) of the average funnellng capacity $\langle f(k) \rangle$ against
degree $k$ for different $\gamma$. 
This data also show that there is no simple algebraic relation between
$\langle f(k) \rangle$ and the degree as has been noted for the betweenness 
centrality showing clearly  that betweenness centrality and funnelling are
not trivially related.

\begin{center}
\begin{figure}
\includegraphics[clip,width= 5.0cm, angle=270]{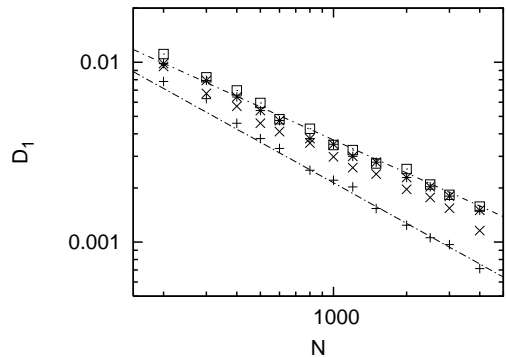}
\caption{ The variation of $D_1$ against $N$ is shown 
for SFN with different $\gamma$ ($\gamma=2.0,2.2,2.4$
and $2.8$). At 
higher values of $\gamma$, the exponents are equal to $\sim 0.60$, while 
for $\gamma=2.0$ it is $ \sim 0.75$.}
\label{funfig4}
\end{figure}
\end{center}

Plotting $D_1$ against $N$, we show that  funnelling  
disappears in the thermodynamic limit for all $\gamma$ (Fig. \ref{funfig4}). 
$D_1$ in fact follows a power law decay with $N$; $D_1 \propto N^{-\rho}$.
For  $\gamma=2$, $\rho \simeq 0.75$ and  decreases from this value as $\gamma$
is increased; beyond $\gamma \sim 2.2$, $\rho = 0.60 \pm 0.01 $ (weakly 
dependent 
on $\gamma$). This  indicates  that the funnelling capacity has universal
behaviour for higher values of $\gamma$. 
Interestingly, the exponent is larger  for smaller values of $\gamma$, i.e., 
  when the 
number of hubs is large.  
This is  consistent with the fact that 
in such a situation, there are multiple routes available for a 
message to reach the targeti, thereby making the 
funnelling capacity  
   of  nodes lesser.   

\begin{center}
\begin{figure}
\includegraphics[clip,width= 5.0cm, angle=270]{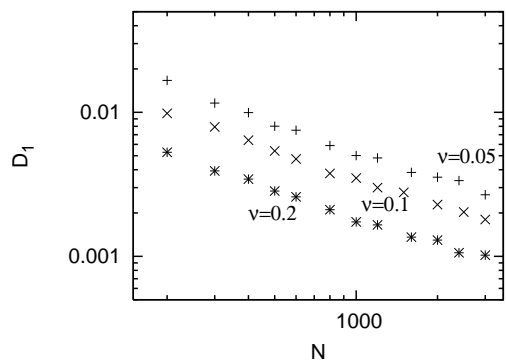}
\caption{ The variation of $D_1$ against $N$ is shown for SFN with  
$\gamma=2.4$ 
and different values of $\nu$ showing that the exponents are not 
dependent on $\nu$.}
\label{funfig5}
\end{figure}
\end{center}

We have also studied   $D_1$ as a function of $N$ for different $\nu$ and find that 
power laws are obeyed for each $\nu$ (Fig. \ref{funfig5}) with the exponent equal to $0.60 \pm 0.01$ in each case. 
The magnitude of $D_1$ decreases linearly with the number of search, which is also
easy to understand (e.g., if there is only one searching process, all the
nodes which take part in this search have funnelling capacity equal to 1,
the maximum possible value).

%Plotted for network of different sizes, we find however, that the value of $D(f)$ decreases
%for increased $N$ indicating that for a very large population, there
%will not exist any funnelling even below $\gamma_c$. 

\subsection {Tunable degree based algorithm}

Next we study the funnelling distributions  on random  scale-free networks with
a tunable 
degree based algorithm.
Precisely, here the search has a preferential algorithm.
During the search, if one of the neighbours of the messenger node
happens to be the target
itself,  the message will be sent to the target. If not, then
the $ith$ neighbour  will  receive the message with a probability
 $\Pi_i$, where
\begin{equation}
\Pi_i \propto k_i^{\lambda}.
\label{lambda}
\end{equation}
Thus here the algorithm can be  extended from 
a random search (RS) ($\lambda =0$) to a highest degree search (HDS) ($\lambda \to \infty$)   scheme as described in
\cite{sen1}. The rule that a message cannot be passed on to the same node twice is still applied.
The essential difference between the present algorithm and the HDS is while the tunable degree based algorithm is stochastic, the HDS is deterministic. 

\begin{center}
\begin{figure}
\includegraphics[clip,width= 5.0cm, angle=270]{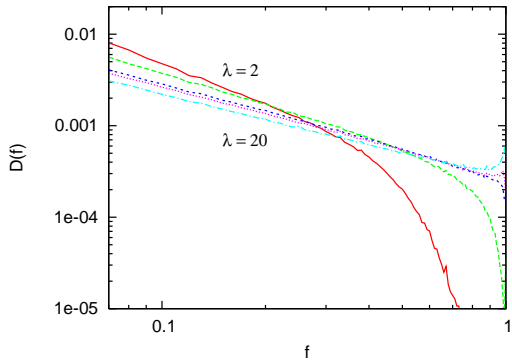}
\caption{ The distribution $D(f)$ is shown for a scale free  network
with $\gamma=2.0$ with different algorithms corresponding to
different values of $\lambda$ (eq (\ref{lambda})).}
\label{alter}
\end{figure}
\end{center}

We have taken scale free networks with $\gamma< \gamma^*$ such that we know that
in the limit $\lambda \to \infty$, it does show a funnelling effect (i.e., power
law degree distribution up to $f=1$).
We find that for small $\lambda$ values, $D(f)$ has a fast
decay as $f$ approaches one, while above a certain value of
 $\lambda= \lambda ^*$ it has a power law decay with the upward bend as noticed
for the HDS. For $\lambda < \lambda_c$, there is a power law behaviour
only over
for a finite range of values of $f$.
The value of $\lambda ^*$ depends on $\gamma$, it being higher for higher values of $\gamma$.
We find that $\lambda^*$ is in fact very high ($\sim 10$ for $\gamma=2.0$)
 (Fig. \ref{alter})
and we have checked that at such large values of $\lambda$, essentially
the signal is being passed to the neighbour having the highest degree.
 The power law decay
of $D(f)$ for large $\lambda$  again
occurs with an exponent close to unity, which is to be expected.

\section{Funnelling in stretched exponential network} 

\begin{center}
\begin{figure}
\includegraphics[clip,width= 5.0cm, angle=270]{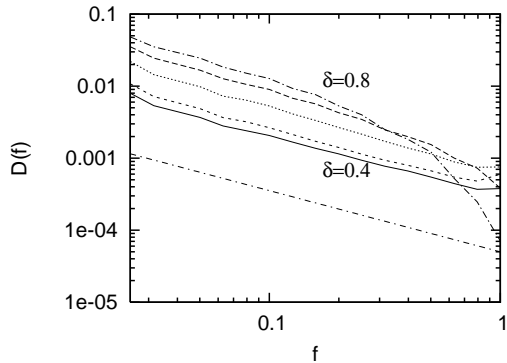}
\caption{ The funnelling distribution $D(f)$ is shown for
stretched exponential  networks with $\delta~=~0.8,~ 0.7, ~0.6,~0.5$ and $0.4$ (from top to bottom) 
with  $\nu = 0.1$. The straight line has slope equal to 0.85.}
\label{stre1}
\end{figure}
\end{center}

There are many real world networks (e.g., social networks)  which do not have a 
scale-free degree distribution. 
 We have therefore considered  networks
in which the degree distribution has a stretched exponential 
distribution: $P(k) \sim \exp(-ck^\delta)$. 
The value of $c$ is unimportant and we set it equal to 1. 
The maximum degree allowed here is $N^{1/2}$.  
Very small values of $\delta$ gives rise to a very highly connected network
which is somewhat unphysical and therefore we  have  taken $\delta > 0.2$.
The funnelling distributions $D(f)$ again shows a power law variation with
a change occurring at $\delta= \delta^*$ (lying between 0.6 and 0.7), above which
the funnelling distribution falls rapidly with $f$. 
The power law exponent is, however, different from that observed in scale
free networks; it has a value close to 0.85.
Once  again we notice that below $\delta^*$, 
$D(f)$ shows a power law decay and a slightly upward bend as $f$ approaches one
(Fig. \ref{stre1}).

\begin{center}
\begin{figure}
\includegraphics[clip,width= 5.0cm, angle=270]{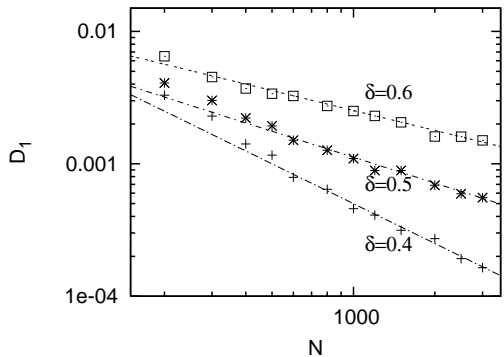}
\caption{ $D_1 = D(f=1)$ against network size $N$ is shown for
stretched exponential  networks $\delta~= ~0.6,~0.5$ and $0.4$ (from top to bottom) indicating  
that the exponents are different as the  straight lines drawn to fit the curves have different
slopes (see text).}
\label{stre2}
\end{figure}
\end{center}

Here too we study the variation of $D_1$ with network sizes and find that  a
power law variation exists, 
  (Fig. \ref{stre2}) however, in this case the exponent is strongly  
dependent on $\delta$. The exponents decrease in magnitude as $\delta$ is
increased, e.g., $\rho \simeq 1.0$ for $\delta = 0.4$ and $\rho \simeq
0.5$ for $\delta = 0.6$. 
Once again we  note that as in the case of scale free networks, 
the exponent for the smaller value of $\delta$ is
higher, when the number of highly connected nodes is larger.

%The transition in the stretched exponential network can be explained in this way:
%from the result of HDS applied to scale free networks we found that 
%there is funnelling effect for $\gamma$ close to 2. 
%It is easy to check that 
%$\exp(-x^\delta)$ is either greater or comparable to $x^{-2}$ for 
%the range of values of $x$ considered here as long as $\delta < 0.7$ and 
%thus even in the stretched exponential network, highly connected hubs are present as in the scale free network.  
%Therefore, as the funnelling capacity has a positive correlation with the 
%highly connected nodes, the funnelling effect is present in the stretched 
%exponential
%network for $\delta < \delta_c$.
%
%\begin{figure}
%
%\includegraphics[clip,width= 5cm, angle=270]{deg_f.eps}
%%\ includegraphics{funnmost1.eps}
%%\includegraphics{funnmost2.eps}
%\caption{ The  variation of the scaled funnelling capacity $f^\prime=  \langle f(k)\rangle (\Delta k)^\gamma$ against $k^\prime = k^\gamma$ is
%shown for scale-free networks with different values of $\gamma$.
%The values of $f^\prime$ for $k^\prime >>1$ coincide 
% for $\gamma=2.0,2.1,2.2$ but   are lesser for $\gamma = 2.3,2.4$ and $2.6.$}
%%               }
%\label{deg-k}
%\end{figure}
%\begin{figure}
%\includegraphics[clip,width= 5cm, angle=270]{strexp.eps}
%%\includegraphics{funnmost1.eps}
%%\includegraphics{funnmost2.eps}
%\caption{ The distribution $D(f)$ is shown for stretched exponential networks for
%different values of $\delta = 0.4,0.6,0.7$ and $0.8.$}
%%               }
%\label{stretch}
%\end{figure}
%

\section{Funnelling in a  correlated network}

Funnelling is an important issue in social searches and 
therefore  we have considered in this section a network which is not entirely 
random in the sense that there is a correlation between nodes. 
The nodes, in reality, have many characteristic features (other than the degree)
which seriously affect the searching process \cite{dodds,watts-search}. 
In a very simplified picture, we consider only one such   characteristic
which we call the similarity factor $\xi$ of the individuals, $\xi$ 
 varying between 0 and 1 randomly. 
 Since we have actually tried  to
simulate a social network, the degree 
distribution is taken to be  $ P(k) \propto \exp(-k)$. 
However, while constructing the network with such a distribution,
the bonding between two nodes
is now made according to the probability
\begin{equation}
{\cal{P}}_{i,j} \propto |\xi_i -\xi_j|^{-\alpha},
\label{alpha}
\end{equation}
such that for positive values of 
 $\alpha$, similar nodes will have more connection probability.
As an example, $\xi$  may simply denote the geographical position of a  
node. The algorithm used is a greedy one: while searching, a node here sends the signal to a 
node with the similarity characteristic closest to that of the 
target node.

\begin{center}
\begin{figure}
\includegraphics[clip,width= 5.0cm,angle=270]{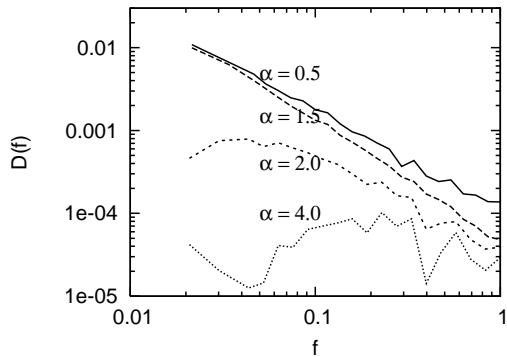}
\caption{ The distribution $D(f)$ is shown for a   network
where the linkings depend on a similarity factor 
for different values of $\alpha$ (eq (\ref{alpha})).}
\label{social}
\end{figure}
\end{center}

Searching on a generalised class of stretched exponential  networks, (i.e.,
those with degree distribution $\exp(-k^\delta))$ with  similarity
dependent  connections has been recently considered \cite{sen3}  and it has been
observed that the
 success rate of searching is drastically reduced as $\delta$ is increased. 
Using  a similarity
based search algorithm and varying the parameter   $\alpha$, 
it was observed that 
the best searchability occurs at     values of $\alpha$ close to 
 1.5
for any $\delta$. For both higher and lower 
 values of $\alpha$ the searchability deteriorates. At large $\alpha$,  the
network is highly clustered in the sense that nodes which have comparable
values of $\xi$ happen to have a strong bonding and since the target node
is randomly selected the success rate falls.
 On the other hand,  
at small $\alpha$, the nodes are highly uncorrelated which makes searching based on similarity ineffective.

As far as funnelling is concerned, we find some intriguing results
which show that searchability and funnelling properties 
are not simply related. Here we have a considered a degree distribution
$\propto \exp(-k)$ where the searchability is rather poor \cite{sen3}.
Still, for small values of $\alpha$, 
there is a power law variation of $D(f)$ with an exponent close to 1.5 (weakly 
dependent on $\alpha$). 
For higher values of $\alpha$, 
large fluctuations occur, 
the power law behaviour is lost and the distribution
tends to become flat (see Fig. \ref{social}). 
 The change in behaviour of the distribution might suggest that a  phase transition is occurring here, 
but we would not like to conclude anything as the 
fluctuations are too large to comment.
% The success rate being small here, good statistics is hard 
%to obtain such that the  fluctuation in the data  is appreciable. On the
%other hand, for small $\alpha$, there is a power law variation in
%$D(f)$ with an exponent which is weakly dependent on $\alpha$.

In order to understand the above results, we first discuss the case
 $\alpha = 0$ which corresponds to a network
without any correlation.   This is simply an exponential 
network where
there is no funnelling effect with HDS as noted earlier.
 But with the present algorithm, we 
find that there is indeed a power law variation of $D(f)$. 
What could be the reason for this?
It appears that since the nodes are uncorrelated, the target node 
is connected to nodes 
with arbitrary similarity factors. However not all of them will 
take part equally in the search process 
because of the algorithm and successful paths will be mostly through a few nodes making it
possible the existence of a few nodes with large funnelling capacity.  

Apparently, the funnelling effect diminishes with correlations as the
present results suggest.
In fact, with $\alpha \neq 0$, when we have a correlated network,
there can be several nodes to which the signal can be passed which
are equally `distant' from the target. This effectively makes the 
funnelling capacity of individual nodes lesser which is reflected 
in the distribution.
For very high $\alpha$, this effect is enhanced to a large extent making
the distribution nearly flat.

\section {Discussions and concluding remarks}

In this paper
 we have, for the first time to our knowledge, attempted a
quantitative study of the phenomenon of funnelling relevant to  
search or navigation on a network.
First we have proposed a definition of funnelling capacity $f$ of a node and thereafter estimated the
 funnelling distribution $D(f)$.
The point $f=1$ has been treated specially as a non-zero value of $D_1=D(f=1)$
would indicate funnelling is indeed occurring. Our studies on scale-free and stretched exponential 
networks have shown that 
funnelling will not survive for infinite networks and also decrease if the 
number of searches is increased. However, we have obtained power law decay behaviour for both $D(f)$ versus $f$ and $D_1$ versus $N$ 
variations. 
The exponent for $D(f)$ is  different for different networks. 
In case of the scale free network, we note that the exponent $\rho$
obtained from the $D_1 $ versus $N$ plots  
is weakly dependent on $\gamma$ for   $2.2< \gamma< 3.0$  while in stretched exponential 
networks, it is non-universal. 
 We have also used different algorithms in the different networks, e.g., degree based algorithms
for networks which are uncorrelated and similarity based algorithm on correlated networks; the results show that the algorithm  seriously
affects the funnelling distribution.

To show how the algorithm can affect the funnelling capacity 
one can take the example of a simple hypothetical network. Suppose the  
network has uniform degree $k=l$ with a tree structure. Thus the successful
paths to the target will flow with equal probability 
through its  
$l$ neighbouring nodes when 
the algorithm does not take into account any correlations (note that 
 the HDS and the random search  are identical in this case). 
Thus these $l$ nodes will take part in $1/l$ fraction of searches, their 
$l^2$ neighbours also take part in $1/l^2$ searches and so on making $D(f) \propto 1/f$. However, if now an algorithm based on correlations is
used, the message flow will no longer be uniform and the distribution
$D(f)$ will be quite different.
In fact we have obtained exponents for $D(f)$ which are different but of the order of unity  in the different networks.

%Making $\lambda$ non-zero, we have a correlated model and we find that
%the $D(f)$ loses its power law decay characteristic as $\lambda$ is increased
%indicating that for correlated networks, there will be no funnelling effects 
%with the present  algorithm.

To explain the result obtained in  \cite{dodds} that there is no
funnelling effect in a social search, one can argue on the basis
of the present results that this is due to the fact that the human network
is far from random and has quite strong correlations.   
On the other hand, in the earlier experiments \cite{milgram}, funnelling was observed 
since the number of searches conducted (compared to the network size) was
very small.

{\bf{Acknowledgment}}: Financial support from CSIR grant no. 3(1029)/05-EMR-II
  and UGC UPE (computational group) is acknowledged. Hospitality at Abdus Salam ICTP, Trieste, where part of the
work was done, is also acknowledged.

\end{document}